\documentclass[final,3p,times,twocolumn]{elsarticle}

\usepackage{epsfig}

\usepackage{amssymb}
\usepackage{amsthm}
\usepackage{amsmath}
\usepackage{graphicx}
\usepackage{dcolumn}
\usepackage{bm}
\usepackage{textcomp}
\usepackage[centerlast]{caption}
\usepackage{subcaption}
 \usepackage{lineno}

\journal{Physics Letters A}
\begin{document}

\begin{frontmatter}


\title{Interaction of Love waves with coupled cavity modes in a 2D holey phononic crystal}
\author[1]{Yuxin Liu\corref{yuxin}}
 \ead{yuxin.liu@phd.ec-lille.fr}
 \cortext[yuxin]{Corresponding author}

\author[1]{Abdelkrim Talbi}%

\author[1]{Bahram Djafari-Rouhani}
 \address[1]{Univ. Lille, CNRS, Centrale Lille, ISEN, Univ. Valenciennes, UMR 8520 - IEMN \& LIA LICS/LEMAC, F-59000 Lille, France}

\author[2]{El Houssaine El Boudouti}%
\address[2]{LPMR, Department of Physics, Faculty of sciences, University Mohammed I, 60000 Oujda, Morocco}

\author[3,4]{Lucie Drbohlavov\'{a}}
\author[3,4]{Vincent Mortet}

\address[3]{Institute of Physics, Academy of Sciences Czech Republic v.v.i, Prague 8, Czech Republic}
\address[4]{Czech Technical University, Faculty of Biomedical Engineering, Kladno, Czech Republic}

\author[1]{Olivier Bou Matar}

\author[1]{Philippe Pernod}

\begin{abstract}
The interaction of Love waves with square array of pillars deposited on a cavity defined in a 2D holey phononic crystal is numerically investigated using Finite Element Method. 
First, the existence of SH surface modes is demonstrated separately for phononic crystals that consist of square arrayed holes, or rectangular arrayed Ni pillars, respectively in, or on, a \textrm{SiO$_{2}$} film deposited on a ST-cut quartz substrate. 
The coupling between SH modes and torsional mode in pillars induces a transmission dip that occurs at a frequency located in the range of the band-gap of the holey phononic crystal. 
Second, a cavity is constructed by removing lines of holes in the holey phononic crystal and results in a transmission peak that matches the dip. The optimal geometrical parameters enable us to create a coupling of the cavity mode and the localized pillar mode by introducing lines of pillars into the cavity, which significantly improved the efficiency of the cavity without increasing the crystal size. The obtained results will pave the way to implement advanced designs of high-performance electroacoustic sensors based on coupling modes in phononic crystals. 
\end{abstract}


\end{frontmatter}


\section{Introduction}
Phononic crystals (PnCs) have received increasing attention in the last two decades as an analogue of photonic crystals (PtCs). \cite{martinez-sala_sound_1995,liu_locally_2000,khelif_trapping_2003,pennec_two-dimensional_2010,zhu_holey-structured_2011,yang_extreme_2014,page_focusing_2016,xu_implementation_2018} Coupling modes have been investigated in both PnCs \cite{khelif_transmittivity_2002,pennec_acoustic_2005,sun_analyses_2005,wang_coupling_2014,korovin_strong_2017,oudich_rayleigh_2018} and PtCs \cite{ota_vaccum_2009,wang_photonic_2011,ohta_strong_2011,dundar_multimodal_2012,li_achieving_2013}, exhibiting features such as energy transferring \cite{pennec_acoustic_2005,sun_analyses_2005,wang_photonic_2011}, absorption \cite{khelif_transmittivity_2002,wang_coupling_2014}, wave confinement \cite{wang_photonic_2011} and frequency modulation \cite{korovin_strong_2017}. 
Most of the research is based on the cavity/waveguide \cite{pennec_acoustic_2005,wang_photonic_2011,korovin_strong_2017}, cavity/cavity \cite{dundar_multimodal_2012,korovin_strong_2017} or waveguide/ waveguide \cite{sun_analyses_2005} systems.
Wang et al. \cite{wang_coupling_2014} demonstrated the transmission cancellations due to the coupling modes of a 1D locally resonant PnC.  
Ohta et al. \cite{ohta_strong_2011} studied the coupling between a PtC nanobeam cavity and a quantum dot. The emission peak of the quantum dot could be tuned by temperature to cross the cavity mode which splits into two peaks. In recent years, peak splitting phenomena have also been discussed in the coupling of two PnC cavities \cite{korovin_strong_2017}, giving rise to a capacity of frequency modulation as well as improved quality factors. 
We believe that it is conceivable to investigate similar effects in different PnC devices.

Recently, cavity modes for Love waves, which are shear horizontal (SH) surface acoustic waves (SAW), are demonstrated in a holey PnC
(h-PnC) that consist of square arrayed holes in a silica (\textrm{SiO$_{2}$}) film deposited on a ST-cut quartz \cite{liu_highly_2018}. The cavity mode can be confined by increasing the crystal size on each side of the cavity. However, this will decrease the excitation efficiency and the quality factor only reaches 1100 with an obvious decrease in transmission due to energy loss. 
This letter investigates the interaction of Love waves with a Ni pillar-introduced cavity in the h-PnC. Love waves are analyzed separately in the h-PnC and the pillared PnC (p-PnC). A mode-coupling induced transmission dip for the p-PnC is found in the band-gap region of the h-PnC. An optimally sized cavity in the h-PnC gives rise to a cavity mode that matches the transmission dip. The cavity mode is coupled to the pillar mode by introducing the pillars into the cavity, which significantly improves the sharpness of the transmission peak without increasing the crystal size while maintaining a high transmission level. These effects are used to design the micro-electromechanical resonators with highly confined cavity modes. The band structures and the transmission spectra are calculated with the finite element method(FEM, COMSOL Multiphysics$^{\circledR}$).

\section{Unit cells resonant properties}
\begin{figure}[h!]
	\centering
	\includegraphics[width=.85\linewidth]{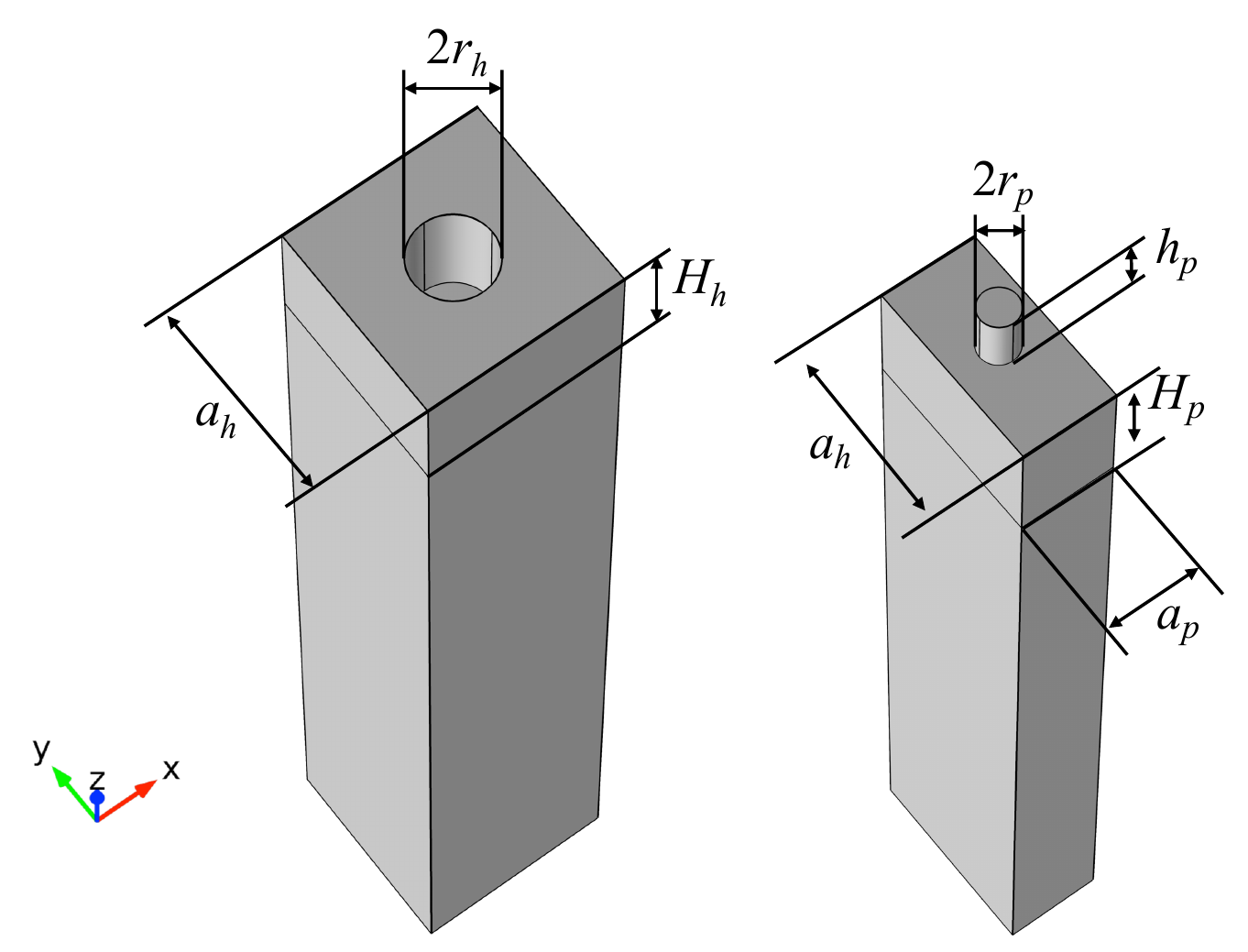}
	\caption{Unit cells of the PnCs with square arrayed cylindrical holes, or rectangular arrayed Ni pillars, respectively in, or on, the silica film deposited on a 90ST-cut quartz. $r_{h}$=0.2$a_{h}$, $r_{p}$=0.2$a_{p}$, $H_{h}$=$H_{p}$=2.4$\mu\textrm{m}$, $h_{p}=0.6a_{p}$, $a_{h}=4\mu\textrm{m}$, $a_{p}=2\mu\textrm{m}$.}
	\label{UC}
\end{figure}
\begin{figure}[]
	\centering
	\includegraphics[width=0.9\linewidth]{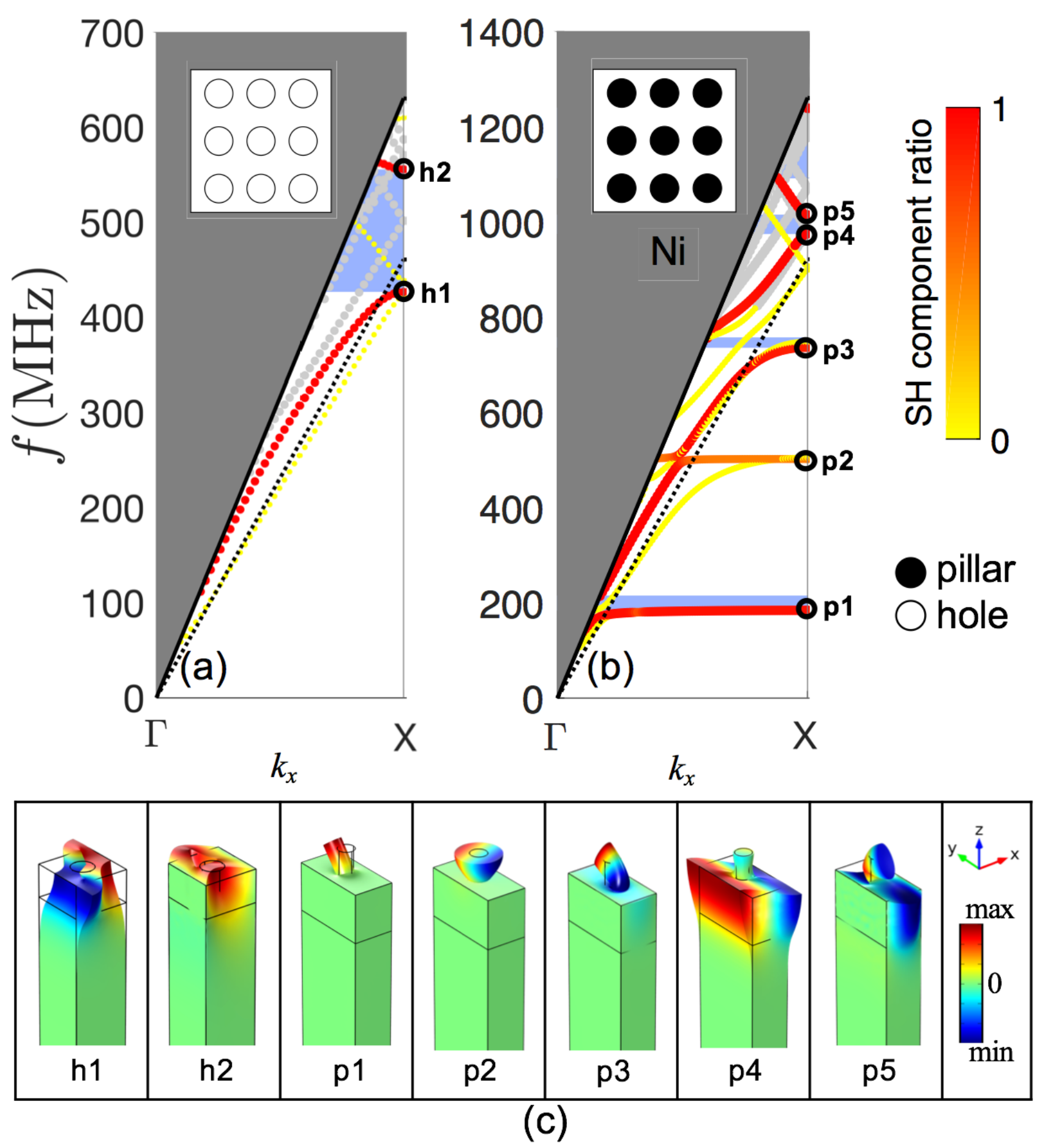}
	\caption{Band structures of the PnCs based on (a) holes and (b) Ni pillars along the $x$ direction. Red-yellow colors denote the SH displacement component ratio. Red indicates SH surface modes. Gray lines denote the modes propagating into the volume. Blue zones are the band gaps in the $x$ direction. The slopes of black solid and dotted lines are the SH velocities in the substrate and SiO$_{2}$, respectively. (c) $u_{y}$ component and deformations of SH surface modes for holey PnC and pillared PnC. $r_{h}=0.2a_{h}$, $r_{p}=0.2a_{p}$, $H_{h}=H_{p}=2.4\mu\textrm{m}$, $h_{p}=0.6a_{p}$, $a_{h}=4\mu\textrm{m}$, $a_{p}=2\mu\textrm{m}$.}
	\label{bandstruct}
\end{figure}

\begin{figure}[]
	\centering
	\includegraphics[width=1\linewidth]{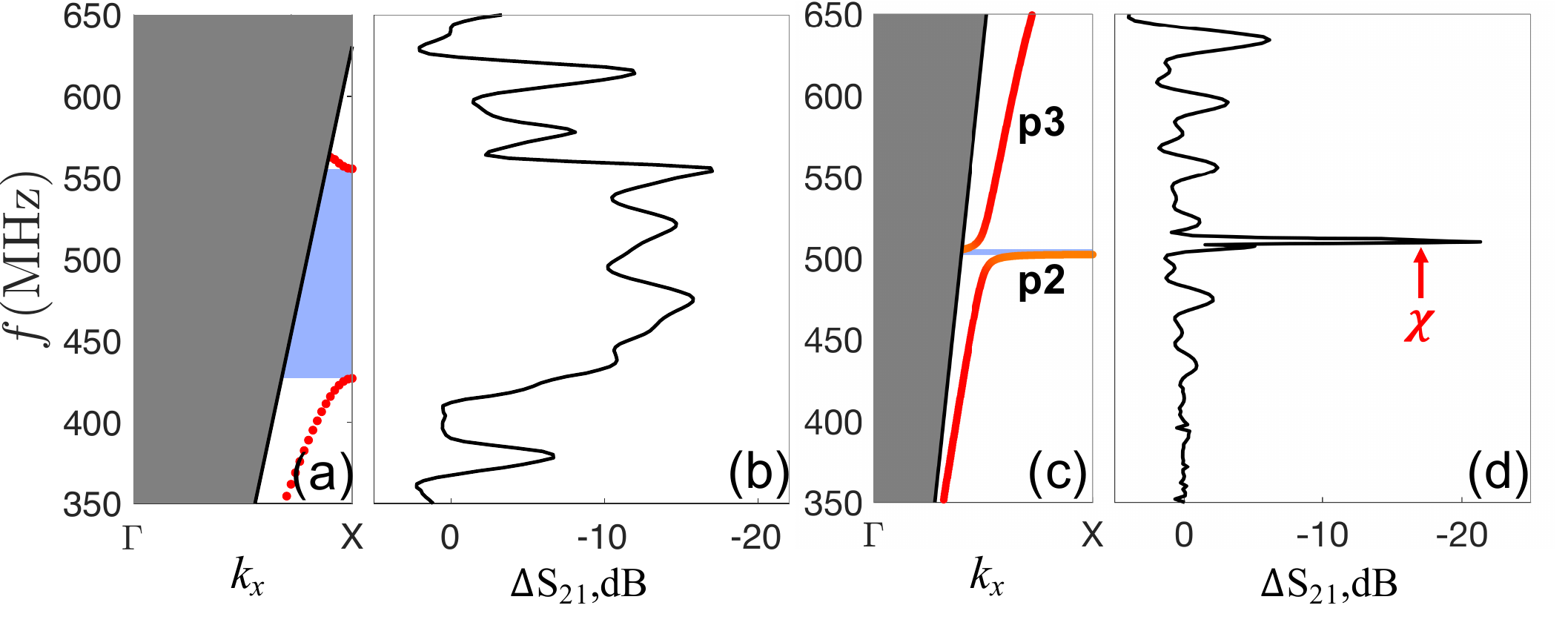}
	\caption{
		(a) and (c) Zoom of the band structures of SH surface modes around the h-PnC band-gap region in the $x$ direction for h-PnC and p-PnC, respectively; (b) and (d) Normalized transmission spectra of SH surface waves propagating through the h-PnC and p-PnC, respectively. 
	}
	\label{TrUC}
\end{figure}

\begin{figure}[hb]
	\centering
	\includegraphics[width=1\linewidth]{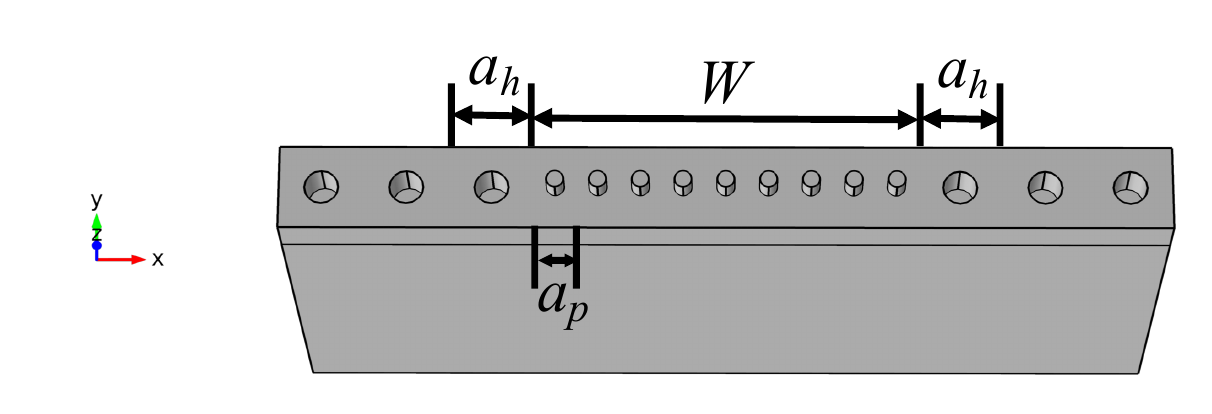}
	\caption{Supercell of pillar-introduced cavity in the h-PnC. $W=4.5a_{h}$, $a_{h}=4\mu\textrm{m}$, $a_{p}=2\mu\textrm{m}$. 
	}
	\label{SC}
\end{figure}

\begin{figure*}[]
	\centering
	\includegraphics[width=1\linewidth]{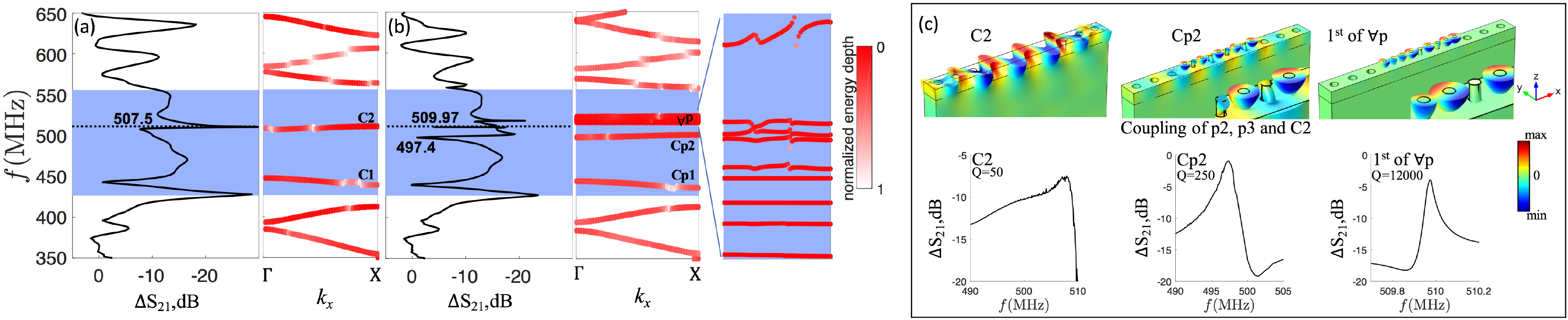}
	\caption{Normalized transmission spectra and band structures of SH surface modes for the cavity-containing h-PnCs (a)without and (b)with pillars. Blue zones are the band-gap region of the perfect h-PnC. Dotted line is the pillar-coupling frequency $f_{\chi}$. (c) Displacement field $u_{y}$ and enlarged transmission peaks of C2, Cp2 and $1^{st}$ of $\forall$p. Insets: zoomed pillars.}
	\label{bsTrW45}
\end{figure*}

The matrix of the PnCs is a $H=2.4\mu\textrm{m}$ height of silica ($\rho=2200$kg/m$^{3}$, $E=70\textrm{GPa}, \nu=0.17$) that covers a $40\mu\textrm{m}$-height 90ST-cut quartz substrate (Euler angles=(0\textdegree, 47.25\textdegree, 90\textdegree), LH 1978 IEEE) which has been rotated 90 degrees around the $z$-axis from the ST-cut quartz, to be able to generate fast SH waves (5000$\textrm{m/s}$) by the electric field.
The shear wave velocity in the silica film is 3438$\textrm{m/s}$, less than that in the 90ST-cut quartz substrate, indicating the existence of Love waves. 
The cylindrical holes (respectively Ni pillars), in (respectively on) the silica film have a radius of $r_{h}=0.8 \mu\textrm{m}$ (respectively $r_{p}=0.4 \mu\textrm{m}$). The period, i.e. lattice constant, of the square arrayed h-PnC (holey PnC) is $a_{h}=4 \mu\textrm{m}$. The periods of the rectangular arrayed p-PnC (pillared PnC) are $a_{h}=4 \mu\textrm{m}$ along the $y$-axis and $a_{p}=2 \mu\textrm{m}$ along the $x$-axis, respectively. The periodicity of p-PnC is chosen to well separate the needed localized mode from the Bragg band gap. The inclusions are chosen because of their strong contrast in density and elastic constants with regard to the matrix. The unit cells of the two PnCs constructed in COMSOL, shown in Fig~\ref{UC}, are employed to calculate the dispersion curves or the band structures. 
Floquet periodic boundary conditions are applied along the $x$ and $y$ directions. The bottoms of the substrates are assumed fixed. Love waves propagate along the $x$-axis (i.e. the $y$-axis of the ST-cut quartz), where Rayleigh waves cannot be generated \cite{liu_design_2014} due to a zero electromechanical coupling factor in the substrate.
The surfaces of the PnCs coincide with the plane $z=0$. The wavelength normalized energy depth (NED)\cite{liu_highly_2018} is calculated to select the surface modes:  
\begin{equation}
\textrm{NED}=\frac{\iiint_{\mathcal{D}}\frac{1}{2}T_{ij}S_{ij}^{*}(-z)dxdydz}{n\lambda\iiint_{\mathcal{D}}\frac{1}{2}T_{ij}S_{ij}^{*}dxdydz} ,
\end{equation}
where $T_{ij}$ is the stress and $S_{ij}$ the strain. The star symbol (*) means the complex conjugate. $-z$ denotes that we integrate in the whole domain of the unit cell $\mathcal{D}$. $\lambda$ is the wavelength. 
For a relatively small $k$ (with wave speed greater than the SH waves velocity of the substrate), $\lambda$ is fixed to $2a$ that is resulting from $k=\frac{\pi}{a}$ and $k=\frac{2\pi}{\lambda}$. $n=1$ for the h-PnC and $n=1.3$ for the p-PnC because of the pillar height, which equals to $0.6a=0.3\lambda$. Surface modes have a NED $<1$. The NED can filter out the bulk modes as well as the plate modes that appear in our finite-depth substrate which is supposed to be semi-infinite for Love waves.

SAW include SH type SAW and Rayleigh type SAW. The SH ratio is calculated to distinguish the displacement components: 
\begin{equation}
\textrm{SH ratio}=\frac{\iiint_{\mathcal{D}}u_{SH}u_{SH}^{*}dxdydz}{\iiint_{\mathcal{D}}(u_{x}u_{x}^{*}+u_{y}u_{y}^{*}+u_{z}u_{z}^{*})dxdydz} ,
\end{equation}
where $u_{x}, u_{y}$ and $u_{z}$ are respectively the displacement along the $x, y, z$ directions. $u_{SH}$ is the SH displacement component that can be expressed as $u_{x}\cos\theta-u_{y}\sin\theta$, which is perpendicular to the wave vector $\bm{k}$. $\theta$ is the angle between $\bm{k}$ and the $y$-axis with $\tan\theta=\frac{k_{x}}{k_{y}}$. 
The band structures of h-PnC and p-PnC in the $x$ direction are shown in Fig~\ref{bandstruct}(a) and (b). The X points are the irreducible Brillouin zone limit of the PnCs in the $x$ direction. Dark gray areas are the radiation zone, where the waves propagate into the volume (the bulk waves). Black lines are the dispersion curves of the SH waves (here the fast shear waves) in the substrate, according to $v=\frac{2\pi f}{k}$. Black dotted lines are the dispersion curves of the shear waves in the SiO$_{2}$ layer. A mode between these two lines is a guided mode in the SiO$_{2}$ layer decaying into the substrate, and a mode below the black dotted line is an evanescent mode or a localized mode in the pillars. The curves in red and yellow denote the surface vibrating modes. Certain modes become gray as they are no more confined to the surface and are filtered out by NED, referred to as leakages. The modes colors are determined by their SH ratio. The red modes have a large SH ratio, indicating the Love modes. Blue zones are the band gaps for SH SAW. For the h-PnC, a Bragg band gap from 426.8 to 555.5 MHz is observed between the two Love modes.
As the wave vector $k_{x}$ increases, it is found for the p-PnC that the modes p2 is coupled to the SH waves around 505MHz, which falls in the band-gap region of the h-PnC, as they approach to each other. 
These two branches exchange their dispersion relations and displacements before they are separated.  
The displacement fields and the deformations of the SH surface modes at point X are shown in Fig~\ref{bandstruct}(c). Due to their large SH component ratio, as well as the exclusive generation of SH waves by the electric field, we only show the transverse component $u_{y}$.
h1 and h2 are guided modes in the h-PnC. h1 is below the dotted line since the holes have a capacity to slow down the Love waves. p1, p2 and p3 are localized pillar modes in the p-PnC while p4, p5 are guided in the SiO$_{2}$ film. 
The torsion mode p2 also has $u_{x}$ component, and is therefore indicated by orange in the band structure.

The transmission spectra is calculated by simulating the same SAW device using our preceding approaches \cite{liu_highly_2018,yankin_finite_2014}, which has been validated by experiments. The model consists of two parts of aluminum inter-digital transducers (IDTs) with the PnC located between IDTs. 
The input IDT is given a $V_{0}=1V$ harmonic voltage signal.
The output is measured by averaging the voltage difference between odd and even electrodes. 
The width of the electrodes of IDTs $L_{IDT}$ is updated for every frequency in the spectrum according to the relation $L_{IDT}=\frac{\lambda}{4}=\frac{v}{4f}$. $v$ is the velocity of Love waves for $H_{h}=0.6a_{h}$, resulting from the basic dispersion relation of Love waves. That is, each frequency corresponds to a single wavelength, and this model simulates the dispersive SAW devices. 
The frequency responses are then normalized by that of the matrix, referred to as normalized or relative transmissions $\Delta S_{21}$. Fig~\ref{TrUC}(b) shows the normalized transmission spectrum of the h-PnC around the band-gap region. The attenuation appears clear and is consistent with the band structure prediction.  Fig~\ref{TrUC}(d) is the normalized transmission spectrum of the p-PnC. The coupling between the torsional mode p2 and SH waves has induced a sharp transmission dip at 508 MHz, denoted as $\chi$. The origin of this dip is a narrow band gap between 507.6 and 510.2 MHz, resulting from the lifting of degeneracy at the crossing point. 
The slight frequency shift is mainly due to the mesh construction of the FEM. 

\section{Supercell and coupling modes}
By removing lines of holes along the $y$ direction in the h-PnC, a cavity is formed perpendicularly to the waves propagation direction. Here we set the cavity width $W$ to $4.5a_{h}$, since a cavity mode is predicted \cite{liu_highly_2018} at almost the same resonant frequency as $f_{\chi}$. The resonator is realized by introducing the p-PnC into the cavity. A supercell containing 1\texttimes ($6+\frac{W}{a_{h}}$) unit cells is constructed, as shown in Fig~\ref{SC}, with periodic boundary conditions applied along the $y$ direction. Since the crystal size only slightly affects the cavity mode frequencies \cite{korovin_strong_2017,liu_highly_2018}, the transmissions are calculated with 4 holes on each side of the cavity.

The Love waves normalized transmission spectra and band structures for the cavity without and with pillars are shown in Fig~\ref{bsTrW45}(a) and (b), respectively. 
On the bare cavity (without pillars), two cavity modes C1 and C2 appear inside the band-gap of the perfect h-PnC, giving rise to two transmission peaks at 438.9 MHz and 507.5 MHz, respectively. The $2^{nd}$ peak matches the transmission dip $\chi$ at 508 MHz that is denoted by the dotted line ($f_{C2}\cong f_{\chi}$). 
After introducing the pillars, the fundamental cavity modes C1 and C2 are coupled to the pillars and become two new modes Cp1 and Cp2, while new pillar modes $\forall p$ appear above. Due to the injected pillars mass, Cp1 at 433.6 MHz has a slight down frequency shift compared with C1, whereas Cp2 has shifted down to 497.4 MHz to show a well separated peak from $\forall p$. The frequency shift of Cp2 results from both the mass loading and the peak splitting effect that
is attributed to a degeneracy of the cavity mode. \cite{korovin_strong_2017}
Meanwhile, the transmission peak of Cp2 is significantly sharper than that of the mode C2, as a result of the matched frequencies between $f_{C2}$ and $f_{p2}$.
The pillar modes $\forall p$ are different combinations of pillars vibrating in mode p2, with the maximum displacements existing in different pillars. The number of pillar modes equals to the number of pillars, i.e., 9 pillar modes with adjacent frequencies around $f_{p2}$, which can be adjusted by the p-PnC parameters. 
The dispersion curves of $\forall p$ are zoomed on the right of Fig~\ref{bsTrW45}(b). The first pillar mode gives rise to a sharp transmission peak at 509.97MHz. This peak originates from the vibration of pillars inserted at the maximum amplitude positions of the cavity mode C2, i.e. the even pillars. Its high intensity in transmission is achieved when we combine all the pillars to create a strong coupling.
The last mode, with a less smoothing and homogeneous dispersion curve, only arouses a small resonance at 517.8 MHz, resulting from the vibration of pillars inserted at the zero amplitude positions of C2, i.e. the odd pillars.
Other pillar modes disappear in the transmission spectrum because they are little coupled to the SH waves.
This transmission spectrum results from a coupling between the p-PnC and the bare cavity-containing h-PnC.
The differences in sharpness of the transmission peaks for modes C2, Cp2 and the $1^{st}$ of $\forall p$ could be explained after observing their displacement fields shown in Fig~\ref{bsTrW45}(c). When Love waves propagate through the cavity, the displacement of C2 is concentrated in the center of the model and is confined in the silica guiding film. The mode Cp2 is actually a coupling of p2, p3 and C2 modes. The pillars vibrating in mode p2 and p3 are alternated along the propagating axis. The two pillar modes are simultaneously coupled to the fundamental cavity mode C2 guiding in the film. Nevertheless, the maximum displacement amplitude is in the pillars, leading to an energy more confined to the surface (a smaller NED). This is proved by the dispersion curve of Cp2 whose slope becomes smaller and red color is more intense than that of C2. 
As for the $1^{st}$ collective pillar mode of $\forall p$, only tiny displacement exists in the guiding layer, leading to the most confined energy and the sharpest transmission peak. Compared with the bare cavity mode confined by increasing the hole number\cite{liu_highly_2018}, this peak remains at a high intensity since the pillar mode decreases the energy loss, and therefore can reach a much higher quality factor.  
The magnification of these three peaks is shown aside. The quality factor (Q) of the mode Cp2 is 5 times greater than that of the mode C2. The Q of the sharpest peak is 240 times greater than that of C2, with a transmission peak at almost the same frequency. This means that the efficiency of the cavity could be significantly enhanced by introducing the optimally sized pillars without increasing the crystal size (the hole number on each side of the cavity). Note that we can create a pillar mode with only one pillar in the cavity but the coupling result depends on the pillar's position with respect to the displacement amplitude of the cavity mode. Additionally, the transmission and quality factor of the mode Cp2 and the $1^{st}$ of $\forall p$ will decrease if we reduce the pillar number, since the coupling between the adjacent pillars will become weaker.

\section{Conclusion}
In summary, we investigated the SH surface modes in two phononic crystals that consist of square/rectangular arrayed holes/Ni pillars in/on a silica film deposited on a 90ST-cut quartz substrate. The h-PnC gives rise to a band gap within which the pillar torsion mode of the p-PnC is coupled to the SH waves, leading to a sharp transmission dip. An optimally sized cavity is introduced into the h-PnC by removing lines of holes and arouses a cavity mode that matches the pillar-coupling frequency. Pillar-containing cavity for Love waves is first demonstrated by introducing the pillars into the cavity. A coupling of the cavity mode to the localized pillar mode is created with significant improvement on the cavity efficiency without increasing the crystal size. The sharpest transmission peak observed at almost the same cavity mode frequency has a 240 times greater Q. This study could be used for potential PnC applications such as filters and bio-sensors.

\label{}
\section*{Acknowledgment}
The authors thank RENATECH, the French national nanofabrication network, the J.E. Purkyně fellowship awarded to V. Mortet by the Czech Academy of Sciences and Centrale Initiative Foundation for their financials supports.




\end{document}